\documentclass[12pt]{article}

\usepackage{gensymb}
\usepackage{url}
\usepackage{amssymb,amsmath, amsfonts}
\usepackage{cases}
\usepackage{latexsym}
\usepackage{graphicx} 
\usepackage{relsize} 
\usepackage{natbib}
\usepackage{rotating}
\usepackage[includeheadfoot,left=1.in,right=1.in,top=1.in,bottom=1.in,bindingoffset=0.in,nohead]{geometry}
\usepackage[onehalfspacing]{setspace}
\usepackage{booktabs}
\usepackage{appendix}
\usepackage[pdfborder={0 0 0}]{hyperref}
\usepackage{float}
\usepackage{xcolor}
\usepackage{caption}
\usepackage{subcaption}
\usepackage{minibox}
\usepackage{longtable}

\begin{document}

\title{Climate Models  Underestimate the\\ Sensitivity of Arctic Sea Ice to Carbon Emissions}

	\author{Francis X. Diebold\\University of Pennsylvania \and Glenn D. Rudebusch \\Brookings Institution \\$~$}
	
	\maketitle
	
\bigskip

\bigskip
	
	\begin{spacing}{1}
		
		\noindent \textbf{Abstract}:
		Arctic sea ice has steadily diminished as atmospheric greenhouse gas concentrations have increased. Using observed data from 1979 to 2019, we estimate a close contemporaneous linear relationship between Arctic sea ice area and cumulative carbon dioxide emissions. For comparison, we provide analogous regression estimates using simulated data from global climate models (drawn from the CMIP5 and CMIP6 model comparison exercises). The carbon sensitivity of Arctic sea ice area is considerably stronger in the observed data than in the climate models. Thus, for a given future emissions path, an ice-free Arctic is likely to occur much earlier than the climate models project. Furthermore, little progress has been made in recent global climate modeling (from CMIP5 to CMIP6) to more accurately match the observed carbon-climate response of Arctic sea ice.

		\thispagestyle{empty}
				
		\bigskip
		
		\bigskip
		
		\bigskip
		
		\bigskip
		
		\bigskip
		
		\noindent {\bf Acknowledgments}:   For comments and/or assistance we thank the Editor and three anonymous referees, Max Goebel, Philippe Goulet Coulombe, Aaron Mora Melendez, Jack Mueller, Gladys Teng, Boyuan Zhang, and the Penn Econometrics Research Group.  The usual disclaimer applies.
		
		\bigskip
				
						
		\bigskip
				
				
		\bigskip
				
		\noindent {\bf Contact}:  fdiebold@sas.upenn.edu; glenn.rudebusch@gmail.com
					
	\end{spacing}
		
	\clearpage
	
%
%

\section{Introduction}

\setcounter{page}{1}
\thispagestyle{empty}

Climate change and rising average surface temperatures are progressing more rapidly in the Arctic than elsewhere. In particular, the loss of Arctic sea ice coverage has been precipitous, especially when measured at the end of the summer melt season. Currently, about half as much of the Arctic ocean is covered by sea ice in September compared to when Arctic ice satellite measurements began 40 years ago (\citealp{NotzStroeve2018}). The swiftly changing Arctic environment is both a stark indicator of climate change and, in turn, a contributing factor affecting the future evolution of the \emph{global} climate system. The dramatic reshaping of the Arctic---with melting sea ice, ice sheets, and permafrost---will have important influences on the pace and extent of  climate change worldwide. For example, with reduced sea ice coverage and more open ocean, less of the sun's radiation is effectively reflected back into space. This reduced sea ice albedo effect promotes increased global temperatures and feeds back to further Arctic melting (\citealp{StroeveNotz2018}).

An invaluable tool for understanding climate dynamics in recent decades has been the evolving collection of large-scale global climate models.  Such models capture the fundamental physical drivers of the earth's climate through a granular, high-frequency accounting of the dynamics of the Earth's atmosphere, oceans, and surface. These structural models have been very useful for a variety of tasks such as uncovering climate variation, determining event and trend climate attribution, and assessing alternative climate scenarios. However, some of the most dramatic Arctic changes evident in the observed data have been poorly captured by large-scale climate models. Notably, these climate models have generally underestimated the amount of lost sea ice in recent decades (\citealp{Stroeve2007}; \citealp{Stroeve2012}; \citealp{Jahn2016};  \citealp{Rosenblum2017}; and \citealp{DRice}).  Such discrepancies imply that the climate models do not yet adequately describe the underlying physical processes and feedback mechanisms in the Arctic. This failure could have far-reaching implications for the performance and predictive ability of the global climate models -- both in the Arctic and elsewhere. 

To better understand the gap between the actual observations and climate model representations of Arctic dynamics, we examine the linear bivariate relationship between sea ice coverage and carbon dioxide (CO$_2$) levels.  This relationship in the observed data has been described by others for both atmospheric CO$_2$ concentration (e.g., \citealp{Johannessen2008}) and cumulative anthropogenic CO$_2$ emissions (e.g., \citealp{NS2016}).  Indeed, the IPCC Sixth Assessment Report \citep{ipccAR62021} summarizes the research literature on this issue by noting that there is ``high confidence'' that satellite-observed Arctic sea ice area is strongly correlated with cumulative CO$_2$ emissions. This strong Arctic sea ice carbon sensitivity---a defining characteristic of the observed data---has been used to assess the Arctic performance of recent vintages of climate models using their simulations conducted for the Coupled Model Intercomparison Project, phases 5 (CMIP5) and 6  (CMIP6). These two vintages are highly-regarded sources for international global climate model simulations. \cite{NS2016} show that most CMIP5 models display a lower sensitivity than the observational record. Similarly, \cite{notz2020arctic} show that most CMIP6 models also fail to simulate the extent of the observed relationship between sea ice and CO$_2$ emissions.

We extend this research and use a more formal statistical approach that regresses Arctic sea ice on cumulative CO$_2$ emissions to assess the congruence of observations and models. Specifically, we examine the strength of the Arctic sea ice carbon sensitivity in observed and model-simulated data to better understand the past and future trajectory of Arctic climate change. This analysis provides a useful characterization of the actual observed data and a straightforward benchmark for assessing the ability of the global climate models to account for and predict Arctic sea ice loss. 

The substantial differences between estimated statistical representations and the CMIP5 or CMIP6 climate models suggest that the climate models do not adequately capture the underlying physical processes and feedback mechanisms in the Arctic.
Of course, the connection between anthropogenic greenhouse gas (GHG) emissions and sea ice coverage is very complex.  Atmospheric GHG affect air and ocean temperature and circulation patterns, cloud cover and albedo, and precipitation -- all with varying seasonality.  As a result, there is still much uncertainty as to why the large-scale climate models fail to capture the extent of the overall downward trend in Arctic sea ice. For example, \cite{Guarinoetal2020} argue that a better representation of the lower surface albedo of summer melt ponds is needed to account for greater incoming shortwave flux. Alternatively, \cite{NS2016} argue that the climate models underestimate the increase in the incoming longwave radiation for a given increase in CO$_2$. Similarly, \cite{NS2016} downplay the role of oceanic heat transport.

  Our paper is related not only to earlier work by others, some of which we have already cited, but also to our own earlier work.   This paper and  \cite{DRice} have some broad similarity but also very important differences. The broad similarity is that both are concerned (in the context of Arctic sea ice) with evaluating the performance of global climate models, by comparing aspects of model simulations to the corresponding aspects of the observational data. The \cite{DRice} evaluation compares climate model forecasts to statistical trend forecasts focusing on projected arrival dates of a near ice-free Arctic (NIFA).\footnote{Also related is \cite{EATV}, which continues with forecasting analyses as in \cite{DRice}, but which is not concerned with evaluation of dynamical climate models. Instead it develops and explores extensions, variations, and robustness checks for statistical forecasting models.} These NIFA arrival dates differ substantially between the statistical representation (early NIFA) and the climate models (late NIFA), but the question remains as to \textit{why} these projections diverge. 
  
In this paper, we  start to address the ``why", using a very different approach that does not focus on forecasting. It is more structural, regressing sea ice on a key science-based covariate, cumulative carbon emissions, rather than on a black-box ``time trend", in keeping with the broad scientific consensus of a linear ice-emissions relationship. Specifically we compare the carbon-sensitivity of sea ice in climate models and in the observational data. Rather than comparing projected NIFA arrival dates, we document that the climate models show insufficient carbon sensitivity. This result calls for a re-examination of the drivers of ice-emissions relationships in dynamic climate models, which the statistical forecasting analysis of \cite{DRice} could not reveal. 
    
We proceed as follows. In section \ref{data}, we characterize Arctic sea ice carbon sensitivity in observed data.  In section \ref{cmip5},  we characterize  Arctic  sea ice carbon sensitivity in leading CMIP5 models, and we compare it to that in the observed data.  In section \ref{cmip6},  we focus on Arctic  sea ice  sensitivity  in CMIP6 models, assessing not only their agreement with the observed data, but also whether their sea ice  sensitivity improves on the earlier-vintage CMIP5 models.  In section  \ref{BC} we examine whether ``bias correction" helps to improve the congruence between data-based and model-based sea ice sensitivity, as is sometimes suggested.   We conclude in section \ref{concl}.

\section{Arctic  Sea Ice Carbon Sensitivity in the  Historical Record} \label{data}

Various researchers -- notably, \cite{Johannessen2008}, \cite{NS2016}, and \cite{StroeveNotz2018} -- have identified a linear empirical relationship between observed Arctic sea ice coverage and atmospheric CO$_2$ concentration or cumulative emissions. This linear relationship, which fits remarkably well in recent decades, can be expressed as
\begin{equation}
  \label{eq:linear}
  ICE_t = \alpha +  \beta \cdot CARBON_t + \epsilon_t,
\end{equation}
where $ICE_t$ is a measure of sea ice coverage, $CARBON_t$ is a measure of accumulated atmospheric CO$_2$ (in this paper we will focus on cumulative emissions), and $\epsilon_t$ represents deviations from the linear fit.\footnote{Our carbon-trend regression is a cointegrating regression if appropriate, but we do not need to take an explicit stand on non-stationarity. In particular, an OLS carbon-trend regression in levels is consistent regardless of whether the trends are deterministic, stochastic (integrated) but not cointegrated, or integrated and cointegrated \citep{SSW1990}. This result allows us to skirt the unit root minefield, which is helpful because our 40 annual sea ice and carbon observations are not very informative in distinguishing unit roots from nearby alternatives.}$^,$\footnote{Other researchers have examined a similar empirical linear relationship between Arctic sea ice and a measure of global temperature, as in \cite{Winton2011} and \cite{Rosenblum2017}.} The regression intercept, $\alpha$, calibrates the average level of sea ice coverage. The slope, $\beta$, provides a broad measure of the climate response of Arctic sea ice. 
We will refer to $\beta$ as the Arctic sea ice sensitivity or carbon sensitivity.  A negative value of $\beta$ captures the diminishing coverage of Arctic sea in response to the greater accumulation of greenhouse gases in the atmosphere. Equation (\ref{eq:linear}) is at the center of our analysis of both the observed historical data and climate model simulations.\footnote{\cite{Matthews2009} consider a similar proportional relationship between global temperatures and cumulative carbon emissions and the associated temperature climate sensitivity.}

%

\begin{figure}[h]
	\caption{Arctic sea ice area and CO$_2$}
	\label{SIAlinear3}
	\begin{center}
		\includegraphics[trim={0mm 5mm 150mm 5mm},clip,scale=.7]{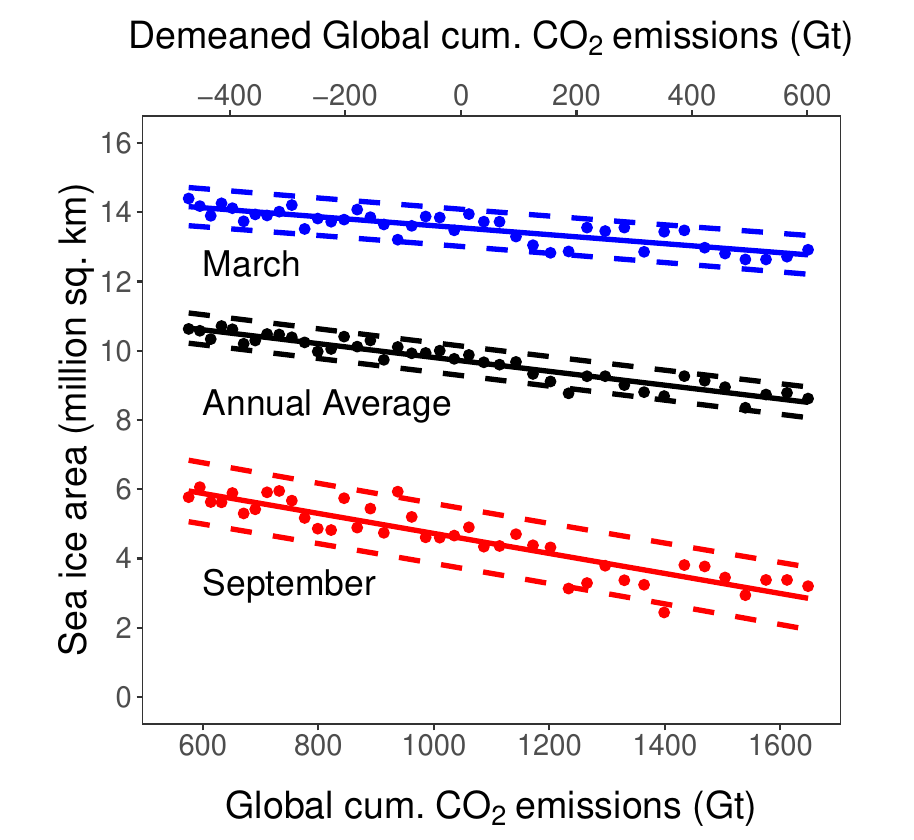}
	\end{center}
	\begin{spacing}{1.0} \footnotesize \noindent Notes: Arctic sea ice area (in million km$^2$) is shown against global cumulative CO$_2$ emissions (in Gt).  March, annual average, and September observations are shown in blue, black, and red, respectively. Dashed lines are  95\% prediction intervals accounting for parameter estimation uncertainty.   Carbon scales are shown in both demeaned (top scale) and non-demeaned (bottom scale) forms.   The sample period is 1979-2019.  See text for details.
	\end{spacing}
\end{figure}

For the observed data, we consider several  of empirical implementations of equation (\ref{eq:linear}) to assess the robustness of the relationship. Some of these variations are shown in Figure \ref{SIAlinear3}, and all data are described in Appendix \ref{datasec}. Arctic sea ice area, $SIA$, is used as a measure of $ICE_t$.  Arctic sea ice coverage has been well measured since the end of 1978 using satellite-based passive microwave sensing. For any polar region divided into a grid of individual cells, the satellite readings provide the fraction of ice surface coverage for each cell. $SIA$ is the sum of the ice-covered areas with at least 15\% ice coverage -- that is, the sum of the fractional cell areas above that minimum.\footnote{In particular, monthly average sea ice data from the National Snow and Ice Data Center (NSIDC) are used, January 1979 - December 2019. These data use the NASA team algorithm to convert microwave brightness readings into ice coverage data. \cite{IcePlus} provide details on data construction and evidence supporting use of NSIDC data.} Similar results are obtained using Arctic sea extent, $SIE$, as a measure of $ICE_t$, as shown in Appendix \ref{appendixb}.

%

In Figure \ref{SIAlinear3} we show $SIA$ against global cumulative emissions as in \cite{NS2016} and \cite{StroeveNotz2018}. We consider three different seasonal  measurements of Arctic sea ice coverage, namely, March (in blue) and September (red) and the annual average (black).  Taken as a whole, the three alternative regressions in Figure \ref{SIAlinear3} display a remarkably consistent linear empirical regularity. In theory, the observed connection between Arctic sea ice and anthropogenic carbon is determined by many different dynamic geophysical channels and feedbacks, including variation in air and water temperature as well as changing surface albedo, cloud cover, wave action, and thermohaline ocean currents. Separately or in combination, these could induce a nonlinear relationship -- or even a tipping point -- between Arctic sea ice and carbon. Instead, for the past few decades, this relationship can be well approximated as linear.\footnote{\cite{NS2016} argue that such linearity can be motivated from a simple conceptual model of the surface energy balance at the sea ice edge.} Of course, when sea ice reaches its zero lower bound, this linear relationship will break down as CO$_2$ levels increase while sea ice -- at least, in late summer and early autumn -- holds steady at its lower bound of zero coverage.\footnote{ \cite{DRice} discuss the lower bound on sea ice and provide an empirical shadow ice modeling strategy to account for it.}  Despite the high likelihood of such future nonlinearities, a linear relationship provides a useful benchmark to capture at a very broad level the current  Arctic sea ice carbon sensitivity.


Table \ref{observed} reports details for the regressions shown in Figure \ref{SIAlinear3}.  The Table \ref{observed} results are based on  demeaned  $CARBON$.   The location shift associated with moving from $CARBON$ to demeaned  $CARBON$ is of course harmless -- just a change of units. Indeed, both $CARBON$ scales are shown simultaneously in Figure \ref{SIAlinear3}, with the demeaned  scale along the top of each panel and the original version at the bottom. However, using demeaned  $CARBON$ aids with the interpretation and comparison of the regression.  In particular, the regression intercept when using demeaned $CARBON$ is the predicted value of $SIA$ at the historical mean of $CARBON$, which is familiar and easily understood.  This contrasts with the regression intercept when using non-demeaned $CARBON$, which is the predicted value of $SIA$ at \textit{zero} $CARBON$, far beyond the range of historical experience. Accordingly, demeaned $CARBON$ data are used from this point onward.

\begin{table}[tb]
	\caption{Regressions of sea ice area on Cumulative CO$_2$ Emissions}
	\label{observed}
	\begin{center}
		\begin{tabular}{lccc}
			\hline\hline
			\\[-2ex]& \multicolumn{3}{c}{Sea Ice Area } \\
			\cline{2-4}
			\\[-2ex]                           & September        & Annual Average        & March           \\
			\hline
			\\[-2ex]Intercept (million km$^2$)              & 4.586            & 9.713                 & 13.553          \\
			& (.066)           & (.033)                & (.041)          \\
			Sensitivity (m$^2$/t CO$_2$)  & -2.891           & -2.000                & -1.299          \\
			& (.209)           & (.103)                & (.130)          \\
			R$^2$                                   & .83              & .90                   & .71             \\
			&                  &                       &                 \\
			H$_0$: Linear regression relationship   & p=.53            & p=.95                 & p=.69           \\
			H$_0$: Stable regression relationship   & p=.45            & p=.28                 & p=.84           \\
			H$_0$: Gaussian regression disturbances & p=.52            & p=.10                 & p=.43           \\
			&                  &                       &                 \\
			NIFA CO$_2$ level                       & 2287             & 5403                  & 10713        \\
			NIFA Year (SSP2-4.5)                    & 2034             &                       &                 \\
			NIFA Year (SSP3-7.0)                    & 2032             & 2080                  &                \\
			\hline
		\end{tabular}
	\end{center}
	\begin{spacing}{1.0} \footnotesize \noindent  Notes: Shown are regression estimates of Arctic sea ice area on demeaned cumulative CO$_2$ emissions since 1850 with coefficient standard errors are in parentheses. Sensitivity or slope coefficients are shown in m$^2$ per ton of CO$_2$ (which is equivalent to thousands km$^2$ per Gt CO$_2$). Also shown are p-values for tests of  three hypotheses: a linear relationship against a quadratic alternative, a linear relationship against a broken linear alternative with mid-sample break point, and Gaussian disturbances against an arbitrary non-Gaussian alternative.  Extrapolated dates of a nearly ice-free Arctic (NIFA) and the associated  CO$_2$ levels assuming SSP2-4.5 and SSP3-7.0 scenario paths are also shown. The sample period is 1979-2019.  
	\end{spacing}
	
\end{table}

The three columns of Table \ref{observed}  are for September, annual, and March $SIA$. The table  reports three types of information.  First, it reports estimates of the intercept and slope coefficients, denoted $\hat{\alpha}$ and $\hat{\beta}$, the coefficient standard errors, and the regression R$^{2}$.  The slope coefficients, ${\beta}$, summarize Arctic sea ice  sensitivity.  There is a clear seasonal pattern, with the September slope steeper than the  annual average, which in turn is steeper than March. Our estimated September sea ice  sensitivity for cumulative carbon emissions is essentially identical to the value in \cite{NS2016}, which was based on a sample from 1953 to 2015.\footnote{\cite{NS2016} also provide an intuitive interpretation of the magnitude of estimated September sea ice  sensitivity, noting that it translates into a loss of approximately 3.0 m$^2$ of September Arctic sea ice per metric ton of CO$_2$ emissions, which allows individuals to easily calculate their own contribution to diminishing sea ice from personal actions.}   The strong linear relationship between the sea ice coverage and carbon forcing -- evident visually in Figure \ref{SIAlinear3} -- is reflected numerically in the small standard errors and high R$^2$'s (more than 80\%) of Table \ref{observed}. 


Second, Table \ref{observed} reports three simple diagnostic test statistics for various aspects of adequacy of the basic regression model  (\ref{eq:linear}).  The first, labeled ``$\rm H_0$: Linear regression relationship" is a t-test of the fitted linear relationship against a nonlinear (quadratic) alternative; that is, a t-test of the coefficient on $CARBON^2$ when added to regression  (\ref{eq:linear}).  The second, labeled ``$\rm H_0$:  Stable regression  relationship," is a \cite{quandt1960tests} F-test   of a stable  linear relationship against the alternative of a broken linear relationship, with a mid-sample break in 2000.  The third, labeled ``$\rm H_0$: Gaussian regression  disturbances," is a \cite{kiefer1983testing} $\chi^2$ test of Gaussian disturbances against an arbitrary non-Gaussian alternative -- effectively a test of skewness=0 and kurtosis=3. In every case -- across all regression variations and hypothesis tests -- there is no evidence that a linear regression is an insufficient representation of the connection between Arctic sea ice and CO$_2$.

As a final statistic of interest, Table \ref{observed} also reports the levels of forcing variables at which the linear regressions predict the effective disappearance of September Arctic sea ice -- a nearly ice-free Arctic (NIFA), which is defined as  only 1 million km$^2$ of sea ice remaining.\footnote{The definition of NIFA follows the usual convention in the literature as the appropriate definition of an effectively ice-free Arctic, reflecting the hypothesized persistence of residual sea ice clinging to northern coastlines despite an open Arctic ocean \citep{EATV}.} The NIFA levels are based on extrapolations of the linear regressions until the effectively ice-free coverage benchmark is reached. A September NIFA is reached with cumulative CO$_2$ emissions of 2287 Gt, which is almost 650 Gt greater than the 2019 observed level of 1648 Gt.\footnote{\cite{NotzStroeve2018} provide a similar benchmark for reaching NIFA.} Table \ref{observed} also shows that this NIFA level of CO$_2$ emissions will be reached in 2032 or 2034 based on the ``Shared Socioeconomic Pathways” SSP3-7.0 or SSP2-4.5. These are two plausible climate scenarios that are widely used as inputs in climate model simulations, especially for assessment reports by the Intergovernmental Panel on Climate Change (IPCC).   Taken together, the results in Table \ref{observed} -- assuming a standard extrapolation of future emissions -- predict an essentially ice-free September Arctic Ocean will likely occur about a decade from now.  This timing is broadly consistent with the statistical projections in \cite{DRice}, \cite{EATV}, and other analyses.\footnote{\cite{DRice} found a slightly increasing rate of decline in Arctic sea ice over the past few decades, which is consistent with a linear relationship between sea ice and CO$_2$, given  the past increasing rate of change in emissions, as discussed in \cite{EATV}.}

\section{CMIP5 Models Have Low Arctic Sea Ice Sensitivity} \label{cmip5}
  
In this section, the observed Arctic sea ice data are compared to individual CMIP5 climate model simulation paths through the  lens of the linear sea ice sensitivity regression  of  $ICE$ on $CARBON$ given in equation (\ref{eq:linear}).    That is, sea ice sensitivity regressions are fit to observed historical data and dynamic model  paths.   In particular, we focus on the September sensitivity regression of $SIA$ on cumulative CO$_2$ emissions, as in \cite{NS2016}, and give attention is both the sensitivity (regression slope) and predicted $ICE$ at mean $CARBON$ (regression intercept).

\begin{figure}[tp]
	\caption{Model and observed sea ice sensitivity,  estimates from single runs of 37 CMIP5 models
	}
	\label{scatterplotannual1run}
	\begin{center}
		Panel A: September estimates
				
		\includegraphics[trim={10mm 0mm 0mm 0mm},clip,scale=.55]{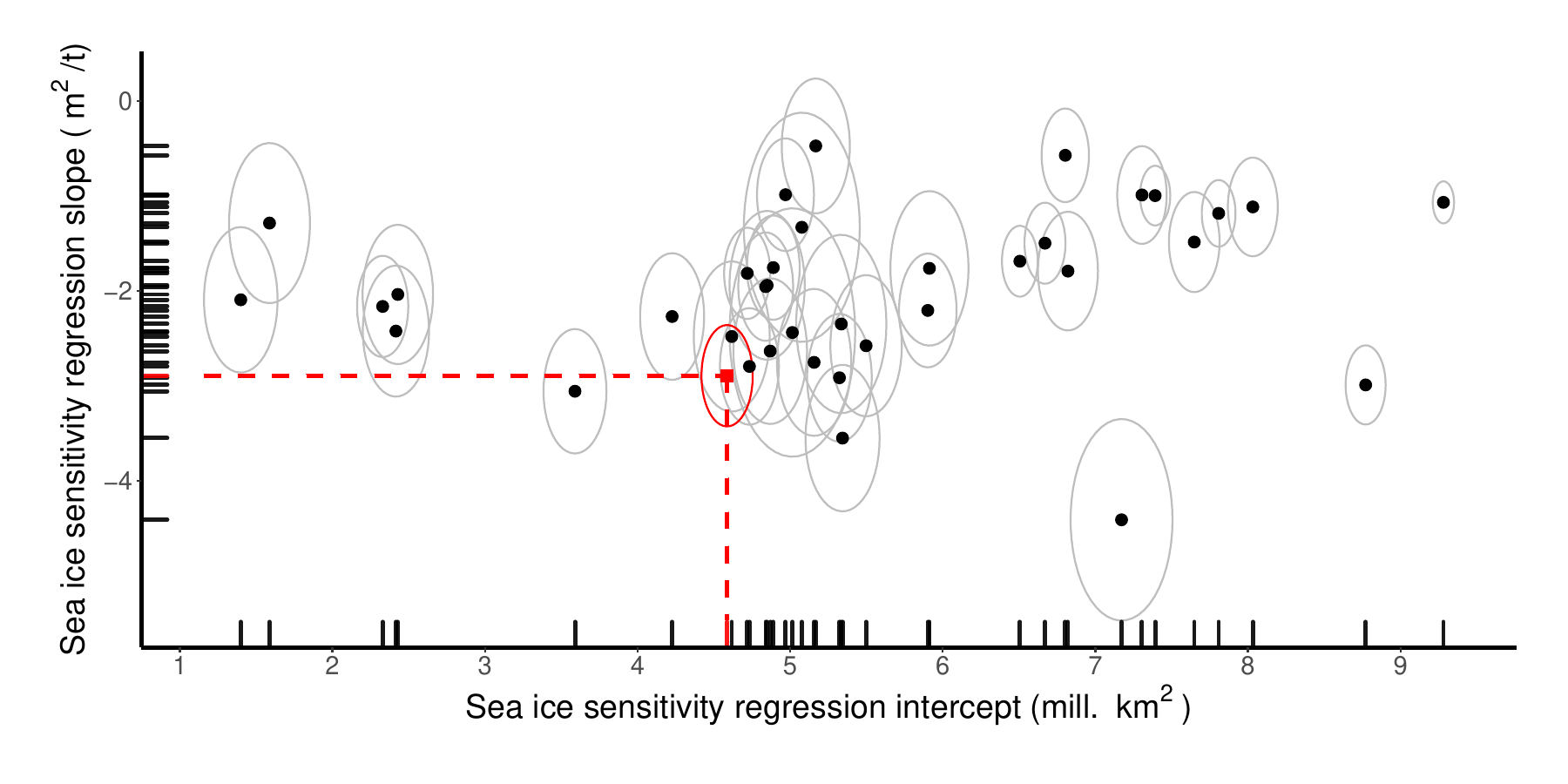}
		
		Panel B: Annual average estimates 
		
		\includegraphics[trim={10mm 10mm 0mm 0mm},clip,scale=.55]{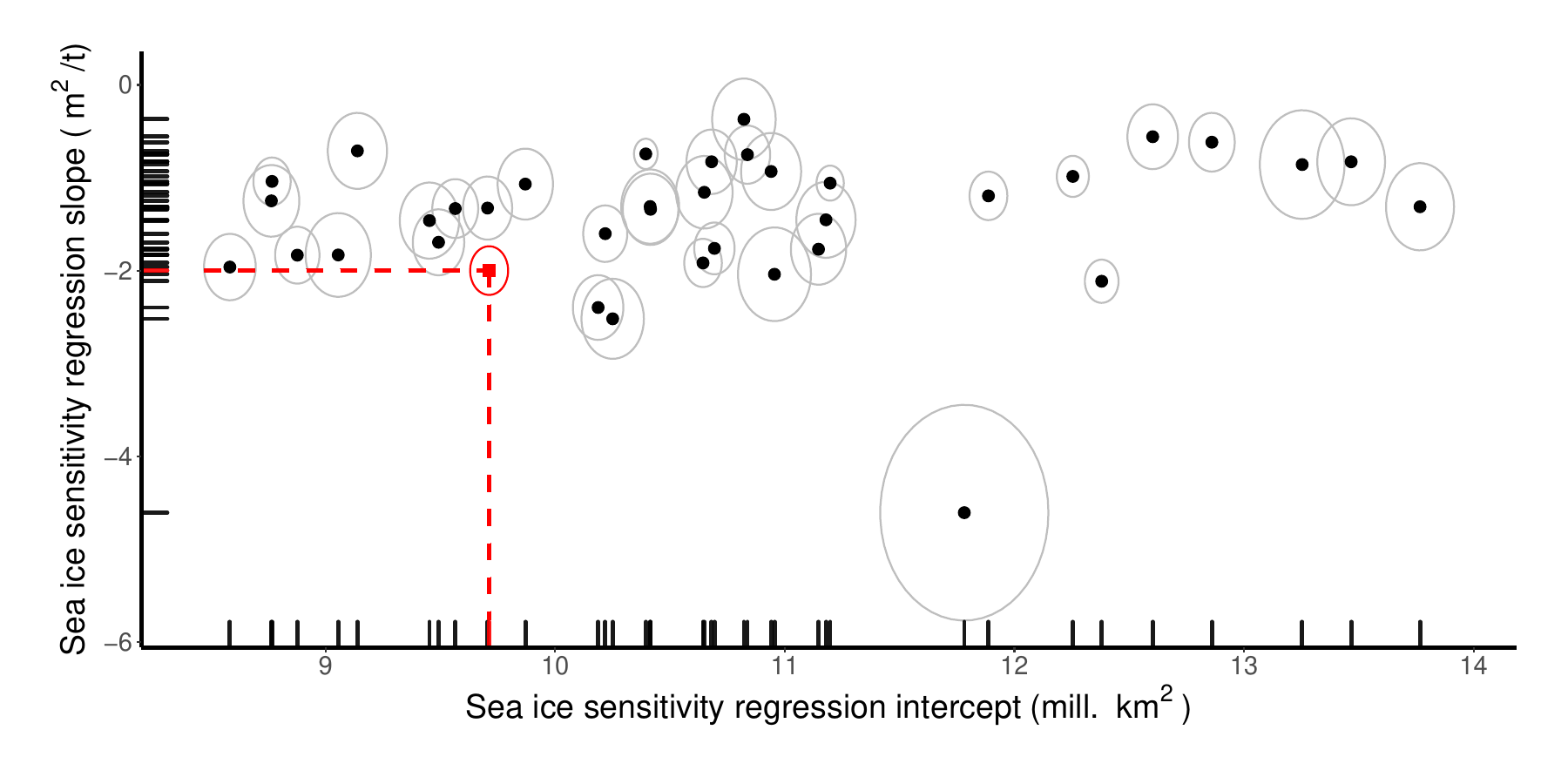}
		
	\end{center}
	
	\begin{spacing}{1.0} \footnotesize \noindent Notes: Sea ice  sensitivity regression intercepts and slopes estimated from  the observed data (1979-2019) are shown as a red square, with a 95\% confidence ellipse.  Sea ice   sensitivity regression intercepts and slopes are shown as black circles, again with  95\% confidence ellipses.  The notches on the axes provide information about  marginal distributions.  See text for details.
	\end{spacing}
\end{figure}

The observational data are summarized by the linear regression estimates given in Table 	\ref{observed}, as discussed earlier. The intercept and slope estimates are shown as a red square in Panel A of Figure \ref{scatterplotannual1run},  together with a 95\% confidence ellipse under normality.\footnote{The confidence ellipses are not tilted with a demeaned carbon series. Since the independent variable is transformed to have zero mean, the sampling uncertainty in the estimation of the slope does not alter estimation of the intercept, and the lack of covariance between the slope and intercept estimate leads to non-tilted ellipses.} These two values (e.g., for September data, $\hat{\alpha}$ = 4.586 10$^6$ km and $\hat{\beta}$ = -2.891 m$^2$/t) will be the key summary statistics for Arctic sea ice climate dynamics that we will use to assess the global climate models.

We first examine the conformity of these estimates with analogous values from 37 different CMIP5 models.  All of these climate models were simulated with a common path of cumulative global CO$_2$ emissions and produced simulated data on Arctic sea ice.  For a single simulation of each model, we use the generated 1979-2019 data sample of emissions and Arctic sea ice area to estimate the Arctic sea ice sensitivity regression.\footnote{The data for the CMIP5 model simulations are detailed in \cite{NS2016} and Appendix \ref{datasec}.  In particular, Figure \ref{scatterplotannual1run} employs the only model run or the first run of any model ensemble under RCP8.5. Detailed CMIP5 regression results are provided in Appendix \ref{appendixc}.}  The resulting 37 linear regression intercept and slope estimates -- one pair for each CMIP5 model -- are shown as black circles with black confidence ellipses obtained from the sampling uncertainty associate with each simulation.

All of the resulting estimates are shown in Figure \ref{scatterplotannual1run}, which  provides a straightforward assessment of whether the black model-based  coefficients match the red data-based coefficients. Panel A of the figure focuses on the September Arctic sea ice sensitivity regressions. Comparing model regression results to results with observed data, only two of the 37 black model parameter point estimates -- those for the CNRM-CM5 and HadGEM2-CC models -- are inside the red data ellipse.  Indeed, for the vast majority of models, the entire black model ellipse has no intersection with the red data ellipse.  Non-overlapping model-based and data-based ellipses indicate that with high probability the population model-based coefficients do not match those governing the observed data, even after accounting for the  uncertainty in both estimates.  



If clear disagreement between models and data is revealed by Panel A of Figure \ref{scatterplotannual1run}, so too is the {nature} of the disagreement. There are three key aspects.   First, the estimated model intercepts have a ``bias problem,'' with most of the model estimates too high. As evidenced by the distributional notches on the horizontal axis, 30 of the 37 black model estimates are to the {right} of the vertical red line. That is, the models tend to be mis-calibrated in terms of the level of sea ice:  \textit{September Arctic sea ice at historical mean carbon is too high in the models}.

Second, the estimated model {slopes} also have a bias problem.  In absolute value, the slope estimates are biased {downward} with 32 out of 37 black model dots above the horizontal red line.  The data-based slope estimate  is approximately -3$\rm m^2$ -- an additional tonne of cumulative emissions causes sea ice loss of 3$\rm m^2$ -- whereas the model-based slope estimates are centered near -2$\rm m^2$.   
That is, the models tend to be mis-calibrated not only in terms of too much sea ice on average but also in terms of the weak absolute response of sea ice to increases in cumulative carbon emissions: \textit{the response of September Arctic sea ice  to increases in carbon is too small in the models}. 

Finally, the figure also makes clear that in addition to being centered in the wrong place, the model-based estimates also have a variance problem.  This is particularly apparent for the distribution of intercepts shown by the notches on the horizontal axis:  \textit{September Arctic sea ice  at historical mean carbon varies disturbingly widely across models}. The intercepts vary from below 2 million km$^2$ to above 9 million km$^2$. The model-based slope estimates have a relatively lower-variance, while the intercept estimates are both high-bias \textit{and} high-variance.

\begin{figure}[t]
	\caption{Model and observed sea ice  sensitivity, ensemble estimates from 12 CMIP5 models}
	\label{scatterplotannual1run3}
	\begin{center}
		
		\includegraphics[trim={0mm 10mm 0mm 0mm},clip,scale=.55]{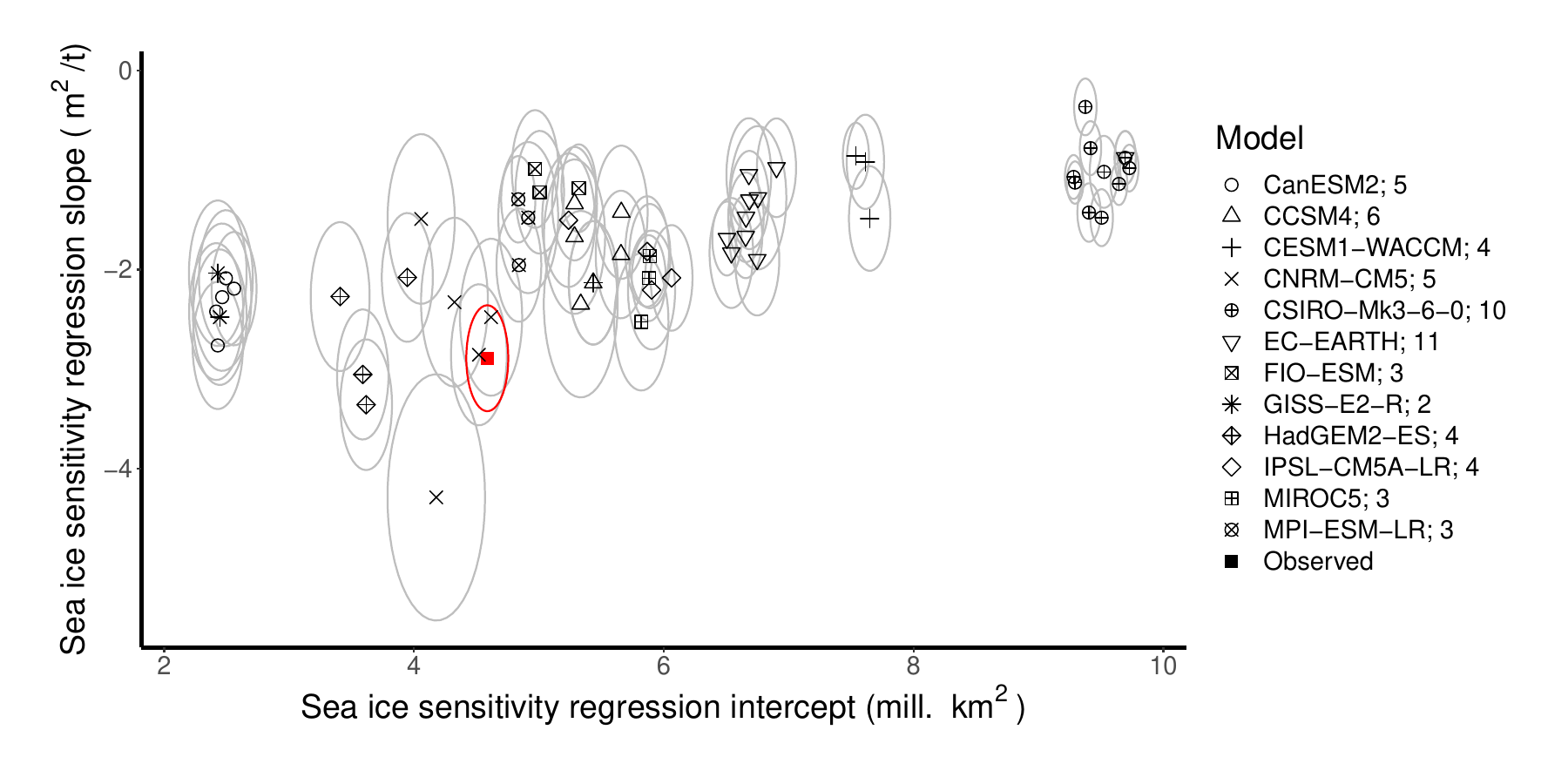}
	\end{center}
	\begin{spacing}{1.0} \footnotesize \noindent Notes: Sea ice   sensitivity regression intercepts and slopes estimated from  the observed data (1979-2019) are shown as a red square, with a 95\% confidence ellipse.  Sea ice   sensitivity regression intercepts and slopes are shown as black circles, again with  95\% confidence ellipses.  Run 5 of EC-EARTH is not shown because it is an extreme outlier with a positive slope of 2.719 and intercept of 4.802.  See text for details.
	\end{spacing}
\end{figure}

Panel B of Figure \ref{scatterplotannual1run} presents a similar comparison but using annual average data, rather than just September.  In principle, the models could perform better at matching Arctic ice dynamics over the whole year and still miss the September lows.  Instead, the intercept bias and intercept variance problems remain, as does the slope bias problem.  
Hence the divergence between models and data is not just a September or a seasonal issue, but the biases in the climate models appear more pervasive. 


Only one run from each CMIP5 model was shown in Figure \ref{scatterplotannual1run}, as in \cite{NS2016}. However, 12 of the 37 models have multiple runs available, i.e., \textit{ensembles} of simulations.  We have replicated our regression analysis on all of the simulations from these models, and in Figure \ref{scatterplotannual1run3}, intercept and slope estimates from these ensembles of model runs are shown for the 12 models.  The estimated intercept-slope pairs from the multiple runs of a single model are all give a unique symbol, and the number of runs in each model's ensemble is reported in the figure key. Taken as a whole, these 50 runs tell the same story as before. Taken as a whole, the multi-run ensembles do just as poor of a job of matching the observational data as regards the connection between Arctic sea ice and carbon emissions.  In particular, (1) Arctic sea ice  at historical mean carbon is too high in most models, (2) the response of Arctic sea ice  to increases in carbon is too weak in most models, and (3) Arctic sea ice  at historical mean carbon varies wildly across models.

The multiple runs in Figure \ref{scatterplotannual1run3} also allow us to address the issue of internal variability. Internal variability refers to variations over time in measures of climate resulting from natural causes. In climate models, internal variability is caused by the climate system’s chaotic nature coupled with slight perturbations in initial conditions.  If, for example, the estimated coefficient pairs for all of the models were distributed evenly throughout Figure \ref{scatterplotannual1run3}, that might suggest that the confidence ellipses based on single simulations generally underestimated the actual climate variation generated by the models.\footnote{\cite{olonscheck2017consistently} argues that internal variability can account for much of the poor fit of climate models to Arctic sea ice.} Instead, all of the coefficient estimates for a given model are clustered together, indicating that our regression estimates are generally robust to internal variability. As in \cite{NS2016}, the Arctic sea ice carbon sensitivity we estimate is based on the average climatic conditions over several decades, which moderates the influence of internal variability to a substantial degree. Indeed, the multiple simulations in Figure \ref{scatterplotannual1run3} tell much the same story as the single runs. Only the same two models -- the CNRM-CM5 and HadGEM2-CC models -- come close to spanning on average the real-world estimated Arctic sea ice sensitivity slope, though both consistently underpredict the intercept and hence the average level of the sea ice. 


\section{CMIP6 Models do not improve on CMIP5} \label{cmip6}

Simulations from the latest generation of climate models -- CMIP6 -- have recently become available. \cite{notz2020arctic} provides an initial overview of these simulations. They conclude that on average the models provide a ``more realistic'' estimate of the sensitivity of September Arctic sea ice area to cumulative carbon emissions than earlier CMIP models.  Here we re-assess that conclusion using the methodology employed above. 

\begin{figure}[tbp]
	\caption{Model and observed September sea ice sensitivity, single simulation runs of 29 CMIP6 models}
	\label{scatterplot1run2}
	\begin{center}
		\includegraphics[trim={0mm 10mm 0mm 8mm},clip,scale=.55]{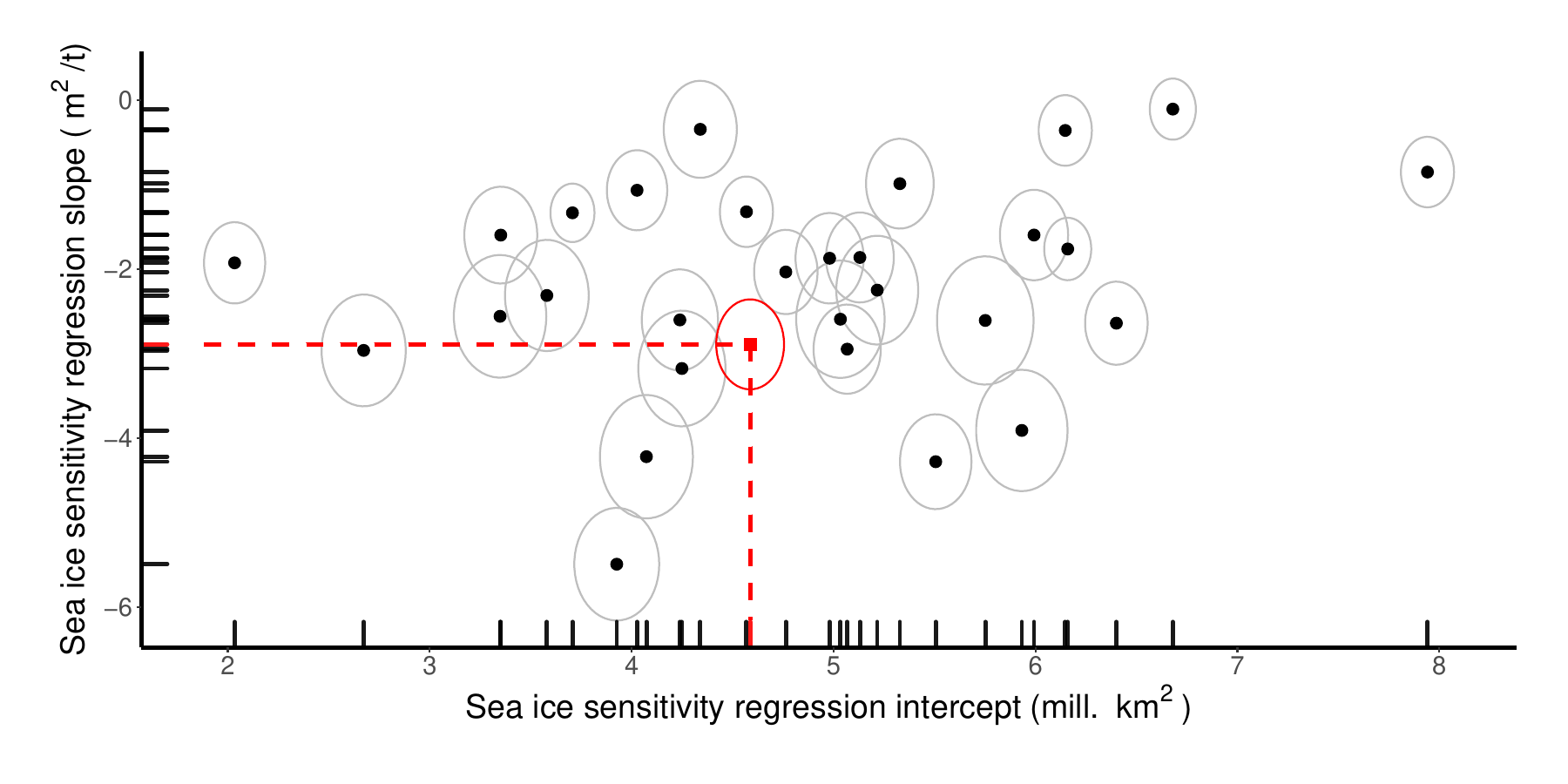}
	\end{center}
	\begin{spacing}{1.0} \footnotesize \noindent Notes: The sea ice sensitivity regression intercept and slope estimated from  the observed September data (1979-2019) are shown as a red square.  Sea ice sensitivity regression intercepts and slopes estimated from single simulation runs of 29 different CMIP6 models are shown as black circles.  The ellipses are 95\% confidence regions for  intercept/slope pairs based on sample variability. 
	\end{spacing}
\end{figure}
 

Recall the key sensitivity visualization of Figure \ref{scatterplotannual1run} (Panel A) for the CMIP5 models, which made clear that Arctic  sea ice sensitivity diverges sharply in the data vs. the CMIP5 models.  The same visualization is shown in Figure \ref{scatterplot1run2}, but for CMIP6 rather than CMIP5 models.  The similarity between Figures \ref{scatterplot1run2} and \ref{scatterplotannual1run} is striking -- the CMIP6 models show little improvement on balance over the CMIP5 models in terms of Arctic sea ice  sensitivity. No model-based coefficient pair falls in the red ellipse. Intercept bias appears to have been reduced, but at the cost of much more notable dispersion across these estimates.  Slope bias seems little reduced, and again, the variance of the estimates has grown.

\begin{figure}[tbph]
	\caption{September sea ice sensitivity results for CMIP5/6 paired models}
	\label{scatterplotCMIP6}
	\begin{center}
		\includegraphics[trim={0mm 6mm 0mm 6mm},clip,scale=0.7]{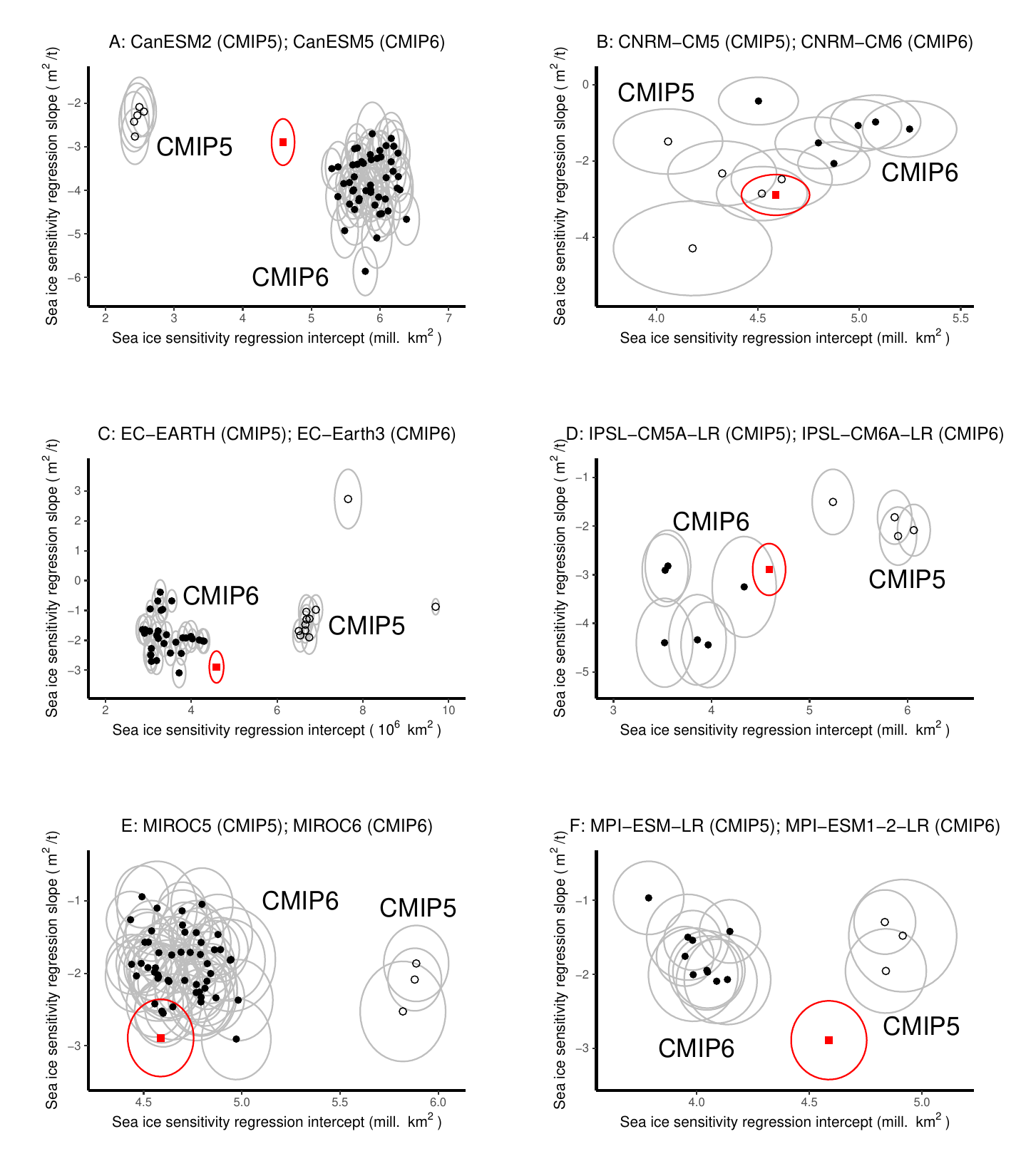}
	\end{center}
	\begin{spacing}{1.0} \footnotesize \noindent Notes: The sea ice sensitivity regression intercept and slope for the observed September data (1979-2019) are denoted by a red square. Similar intercepts and slopes estimated from single simulation runs of two different models are shown as black circles -- CMIP5 as open circles and CMIP6 as solid circles. Each ellipse is a 95\% confidence interval for a coefficient pair based on sample variability. 
	\end{spacing}
\end{figure}

The CMIP6 ensemble results are equally striking, especially when contrasted directly with the CMIP5 ensemble results. As described in the data appendix, we were able to pair CMIP5 and CMIP6 versions of the same basic model in 6 instances in which we also had a reasonable number of simulations for each phase. These basic models are  CanESM, CNRM-CM, EC-EARTH, IPSL-CM, MIROC, and MPI-ESM as shown in Figure \ref{scatterplotCMIP6}. In each panel, the exact versions of the models used are reported in the panel titles. The estimated September Arctic sea ice area response coefficient pairs are given in each panel for the CMIP5 and CMIP6 ensemble simulations. The coefficient pairs from the ensembles of CMIP5 simulations are shown as open circles. and the corresponding pairs from the CMIP6 simulations are shown as solid circles. Of course, given the improvements in modeling and computational speed, there are typically more CMIP6 simulations than CMIP5 simulations in the ensembles. These are detailed in the rightmost two columns of Table \ref{comparison}, so for example, CanESM has 5 runs from CMIP5 and 49 from CMIP6.

On the one hand, as in Panels A, C, D, and E of Figure \ref{scatterplotCMIP6}, the CMIP6 estimates appear to be clustered somewhat closer to observed real-world slope and intercept pair.  On the other hand, however, as for the CMIP5 simulations, there is little or no overlap between the CMIP6 simulations and the observed sensitivity ellipses. There are also notable differences in the CMIP5 and CMIP6  sensitivity ellipses.  Indeed, the \textit{CMIP5 and CMIP6 simulations appear as distinct clusters, overlapping little or not at all with each other or the observed sensitivity ellipse}.  Overcompensation is often apparent, as in Panel A for the CanESM ensembles, whose intercepts desirably increase from CMIP5 to CMIP6, but by too much.

Table \ref{comparison} provides some summary statistics for the panels of Figure \ref{scatterplotCMIP6}.  The first two columns of the table show the distances from the observed intercept and slope estimates ($\alpha^{obs}$ and $\beta^{obs}$) to the median estimates from the CMIP5 and CMIP6 simulations ensemble-by-ensemble ($\bar{\alpha}$ and $\bar{\beta}$). The second two columns show the standard errors of the estimated model-based intercept and slope coefficients across simulation runs in each ensemble ($\sigma(\hat{\alpha}) $ and $\sigma(\hat{\beta}) $). The final two rows give the median of of the six CMIP5 statistics and the median of the six CMIP6 statistics for each column.

The intercept bias in the first column, labeled $\bar{\alpha}-\alpha^{obs}$, where positive values correspond to overestimates of the real-world intercept, which results from simulations with too much sea ice on average.  As shown in the bottom two rows, the CMIP6 models have reduced this bias; however, as shown by the increased intercept standard errors, variability has also jumped. The sea ice sensitivity differences appear in the second column, labeled $\bar{\beta}-\beta^{obs}$, where positive values correspond to insufficient sensitivity.  All CMIP5 models display insufficient sensitivity; the cross-model median  $\bar{\beta}-\beta^{obs}$ is 0.87.  The CMIP6 models fare no better; four of the six remain insufficiently sensitive, and the other two over-compensate and become overly sensitive.  The cross-model CMIP6 median is \textit{worse} than the CMIP5 median, with  $\bar{\beta}-\beta^{obs}= 0.98$. The variance of slope estimates evident in $\sigma(\hat{\beta}) $ has also risen.

\begin{table}[tbph]
	\caption{Comparison of sea ice sensitivity in CMIP5/6 models and observed data}
	\label{comparison}
	\centering
	\begin{tabular}{@{\extracolsep{12pt}}lrrrrc}
		\toprule
		Model (CMIP vintage) & $\bar{\alpha}-\alpha^{obs}$ & $\bar{\beta}-\beta^{obs}$ & $\sigma(\hat{\alpha}) $ & $\sigma(\hat{\beta}) $ & \#Sims \\
		\midrule
		CanESM2 (CMIP5)         & -2.123 & 0.614  & 0.057 & 0.262 & 5   \\
		CanESM5 (CMIP6)         & 1.277  & -0.954 & 0.268 & 0.639 & 49  \\
		&        &        &       &       &     \\
		CNRM-CM5 (CMIP5)        & -0.263 & 0.412  & 0.232 & 1.025 & 5   \\
		CNRM-CM6-1 (CMIP6)      & 0.348  & 1.773  & 0.257 & 0.551 & 6   \\
		&        &        &       &       &     \\
		EC-EARTH (CMIP5)        & 2.099  & 1.591  & 0.932 & 1.293 & 11  \\
		EC-Earth3 (CMIP6)       & -1.290 & 0.971  & 0.420 & 0.637 & 31  \\
		&        &        &       &       &     \\
		IPSL-CM5A-LR (CMIP5)    & 1.298  & 0.941  & 0.363 & 0.311 & 4   \\
		IPSL-CM6A-LR (CMIP6)    & -0.881 & -0.903 & 0.325 & 0.781 & 6   \\
		&        &        &       &       &     \\
		MIROC5 (CMIP5)          & 1.293  & 0.805  & 0.037 & 0.337 & 3   \\
		MIROC6 (CMIP6)          & 0.110  & 0.993  & 0.150 & 0.428 & 50  \\
		&        &        &       &       &     \\
		MPI-ESM-LR (CMIP5)      & 0.253  & 1.414  & 0.045 & 0.341 & 3   \\
		MPI-ESM1-2-LR (CMIP6)   & -0.572 & 1.040  & 0.106 & 0.365 & 10  \\
		\hline \\[-1.5ex]
		Median of CMIP5 models & 0.773  & 0.873 & 0.145 & 0.339 \\
		Median of CMIP6 models & -0.231 & 0.982 & 0.262 & 0.594\\
		\bottomrule
	\end{tabular}
	\begin{spacing}{1.0} \footnotesize \noindent \begin{flushleft} Notes:  Top panel:  For model simulations, differences of the median model simulation intercept and slope from the observed data estimates are shown, respectively, $\bar{\alpha}-\alpha^{obs}$ and  $\bar{\beta}-\beta^{obs}$, 1979-2019.  In addition, standard deviations of the model simulation intercept and slope estimates are shown, respectively, $\sigma(\hat{\alpha}) $ and $\sigma(\hat{\beta})$. Bottom panel: median values of CMIP5 and CMIP6 median-model results.
		\end{flushleft}
	\end{spacing}
\end{table}

\section{Bias Correction Is Not a Solution}  \label{BC}

Our results have illustrated dramatic shortcomings in climate model representations of Arctic sea ice coverage. As noted above, previous research has described related deficiencies in climate model fit and performance.  In response, a literature has developed that ``bias corrects'' climate model simulations. Indeed, bias correction methods of varying sophistication have been used in hundreds of climate change impact studies over the past decade (e.g., \cite{otto2012reconciling}, \cite{turner2013initial}, \cite{melia2015improved}, \cite{ivanov2018climate}, and \cite{kusumastuti2022correcting}). In this section, we argue that such bias corrections are at best a partial fix for climate model projections and that they should not obscure the need for further improvement and progress in climate models.

The bias correction of climate model outputs has been performed using a variety of methods ranging from simple to arcane. The basic goal of a bias correction is to adjust the climate model projection of a particular variable ex post in order to have that projection better match the historical data during some ``calibration sample'' -- and hopefully beyond that as well. For example, a simple ``additive'' bias correction of a series merely adjusts a model simulation by subtracting the difference between the average of that model's ensemble of simulated data for that series and the average of the observed data over the given calibration sample. That is, each individual model simulation is corrected for the average error across all the simulations in the ensemble. To try to account for the effects of the zero lower bound for sea ice indicators, \cite{melia2015improved} also introduce multiplicative bias correction methods based on the ratio of the mean of the model ensemble simulations and the mean of the observed data along with variance corrections.

For the CMIP6 climate model simulations of Arctic sea ice area, we have explored the effects of simple additive bias corrections and the effects of applying the more complicated mean and variance bias correction (MAVRIC) of \cite{melia2015improved}.  For the latter, Figure 6 provides a representative example using the MPI-ESM1-LR model CMIP6 simulations.  Each $SIA$ simulation is adjusted based on the observed bias across the 10 simulations in this model's ensemble from 1979 to 2019. We then use the bias-corrected model simulation $SIA$ paths as data in the sea ice sensitivity regression of equation (\ref{eq:linear}). Figure 6 compares the resulting regression intercept and slope estimates from the original (no bias correction) simulations and the bias-corrected simulations (together with 95\% confidence ellipses). Not surprisingly, the application of the bias correction  re-centers the intercept estimates so that they line up with the historical intercept on average. That is, the bias-corrected simulation intercept estimates do more closely match the observed intercept of the historical series -- the white dots are better centered vertically with red square. However, the bias correction has not tempered the wide variance of intercept estimates for individual simulations. More seriously, the slope estimates remain wildly off the mark from a historical perspective, so the bias correction has done nothing to eliminate the problem of the relative insensitivity of Arctic sea ice conditions to CO$_2$. We obtained similar results with other models and bias correction methods.


\begin{figure}[tbp]
	\caption{September sea ice sensitivity of a bias-corrected CMIP6 model} 
	\begin{center}
		\includegraphics[trim={00mm 0mm 0mm 0mm},clip,scale=.7]{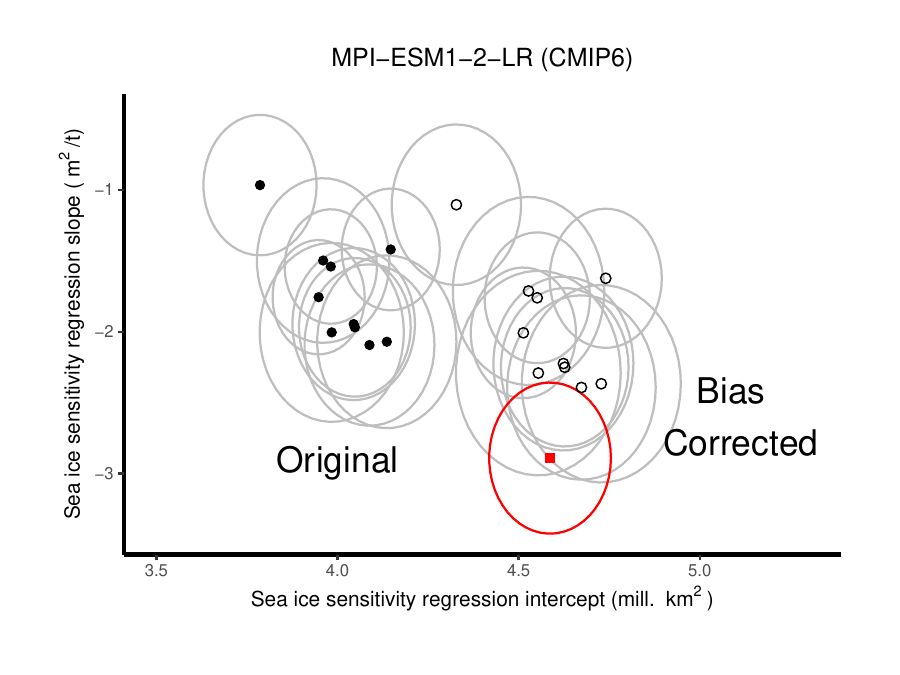}
	\end{center}
	\begin{spacing}{1.0} \footnotesize \noindent Notes: The sea ice sensitivity regression intercept and slope for the observed September data (1979-2019) are denoted by a red square. Similar intercepts and slopes estimated from single simulation runs of the MPI-ESM1-2-LR (CMIP6) ensemble are shown as black circles. Intercept and slope estimates from bias-corrected simulation runs are shown as white circles. Each ellipse is a 95\% confidence interval for a coefficient pair based on sample variability. 
	\end{spacing}
\end{figure}

%
%

Figure 6 illustrates a longstanding trenchant criticism of the use of bias correction in the climate modeling literature (e.g., \cite{ehret2012} and \cite{franccois2020multivariate}). Namely, because bias correction almost invariably focuses on a single variable, it cannot correct a climate model's deficiencies in capturing multivariate interrelationships.  
Furthermore, it should be stressed that even where bias-correction improves the calibration sample fit, there is no reason to assume that the model bias is constant over time, so any bias-corrected projections from a climate model outside the calibration sample are still suspect (\cite{chen2015assessing}).

More broadly, these results raise serious questions about the widespread use of bias-corrected model simulations.  The popular univariate bias corrections correct one physical variable at one location during a given time period and will thus fail to reproduce the inter-variable, spatial, and temporal dependencies of the observations. Furthermore, by ignoring the climate model's cross-variable correlations and intertemporal connections, a bias correction jettisons the very structural connections that might give a climate model a forecasting edge over statistical methods. In addition, working only with bias-corrected output risks losing sight of the size and nature of the underlying biases. Accordingly, bias correction of climate model output as a post-processing step may be better viewed as an ad hoc quick fix. It may be acceptable for some applied work, but a well-founded solution will require better understanding of the sources of any biases, which may provide a path to improve the climate model projections.


\section{Concluding Remarks}  \label{concl}

The substantial differences in Arctic sea ice sensitivity between CMIP5/6 models on the one hand, and the observational record on the other,  imply that the climate models do not adequately capture the underlying physical processes and feedback mechanisms in the Arctic.  Of course, the connection between anthropogenic greenhouse gas (GHG) emissions and sea ice coverage is very complex.  Atmospheric GHG affect air and ocean temperature and circulation patterns, cloud cover and albedo, and precipitation -- all with varying seasonality.  As a result, there is still much uncertainty as to why the large-scale climate models fail to capture the extent of the overall downward trend in Arctic sea ice.  The type of high-level statistical results presented here may be of use in tuning the climate models or identifying problematic aspects of the models.

The fact that climate models  underestimate the sensitivity of arctic sea ice to carbon emissions suggests that   the Arctic Ocean will lose its sea ice or ``turn blue'' at the end of the summer season  sooner than predicted by climate models. This is precisely the conclusion at which \cite{DRice} arrived when comparing climate models models to climate data using very different tools than those developed and  used in this paper. Furthermore, an early arrival of a seasonally ice-free Arctic will likely have important follow-on implications for the pace of climate change around the world.


%
%
%
%
%
%
%
%

\clearpage
\appendix
\appendixpage
\addappheadtotoc
\newcounter{saveeqn}
\setcounter{saveeqn}{\value{section}}
\renewcommand{\theequation}{\mbox{\Alph{saveeqn}.\arabic{equation}}} \setcounter{saveeqn}{1}
\setcounter{equation}{0}

\section{Data}  \label{datasec}

	\subsection{Observed data}
	
	\subsubsection{Arctic sea ice area and extent}
	
	Monthly average area and extent data are from the National Snow and Ice Data Center (\url{https://nsidc.org/data/G02135/versions/3}), measured in millions of square kilometers.  The December 1987 and January 1988 observations are missing because of satellite problems.  We interpolate those observations with fitted values from a regression on trend and monthly dummies estimated using the full data sample.  The reported values for Arctic sea ice area do not include the area near the pole not imaged by the satellite sensor (the ``pole hole"), but, to maintain comparability to climate model output, the pole hole has been added.  Its area is 1.19 million square kilometers from 1979 through August 20, 1987, 0.31 million square kilometers from 21 August 1987 through December 2007, and 0.029 million square kilometers from January 2008 onward.
	  
%
%
	
	\subsubsection{CO$_2$ emissions}
	
	Global annual CO$_2$ emissions data are from the 2020 Global Carbon Budget Project (\url{https://www.icos-cp.eu/science-and-impact/global-carbon-budget/2020}), Fossil emissions excluding carbonation are used, measured in billions of tons of carbon dioxide per year (GtCO$_2$/yr). A cumulative CO$_2$ emissions series is then created, using base year 1850. 

	

\subsection{Climate model data}

%

\subsubsection{CMIP5  climate model data}

The sea ice area data for individual models of phase five of the Coupled Model Intercomparison Project (CMIP5) are based on publicly available output from the replication file for \cite{NS2016}. The historical scenario data range from 1860 to 2005, and the model data for RCP8.5 range from 2006 to 2100.

\subsubsection{CMIP6 climate model data}

The sea ice area data for 29 single run models of phase six of the Coupled Model Intercomparison Project (CMIP6) are taken from the replication file for \cite{bonan2021constraining} (\url{https://zenodo.org/record/5177172}). The historical scenario data range from 1860 to 2005, and the model data for SSP5-8.5 range from 2006 to 2100. 

Data for multi-run CMIP6 models were directly queried from the World Climate Research Program (WCRP) data repository (\url{https://esgf-node.llnl.gov/search/cmip6/}). Data at monthly frequency of sea ice concentration for the historical, SSP1-2.6, SSP2-4.5, SSP3-7.0 and SSP5-8.5 experiments were queried for the CanESM5, MPI-ESM1-2-LR, CNRM-CM6-1, EC-Earth3, IPSL-CM6A-LR and MIROC6 models.

To obtain time series from the queried data, the following steps are necessary: (i) convert raw gridded data in \verb|netcdf4| format to \verb|netcdf-classic|; (ii) use Climate Data Operators (CDO) (\url{https://code.mpimet.mpg.de/projects/cdo}) to convert gridded data from the \verb|netcdf-classic| files into monthly Northern Hemispheric SIA and SIE values; (iii) convert the transformed files into monthly time series. This was done in R with  the resulting series  stored into \verb|csv| files for further use.

The resulting  series contain Northern Hemisphere annual SIA and SIE from 1850-2100 for each model ensemble. In such series the historical range is from January 1850 to December 2014, and the SSP simulations (SSP1-2.6, SSP2-4.5, SSP3-7.0 and SSP5-8.5) start in January 2015.

\section{Sensitivity in Observed Sea Ice Extent} \label{appendixb}

\begin{figure}[tbph]
	\caption{Arctic sea ice extent and CO$_2$}
	\label{SIElinear3}
	\begin{center}
		\includegraphics[trim={0mm 5mm 150mm 5mm},clip,scale=.7]{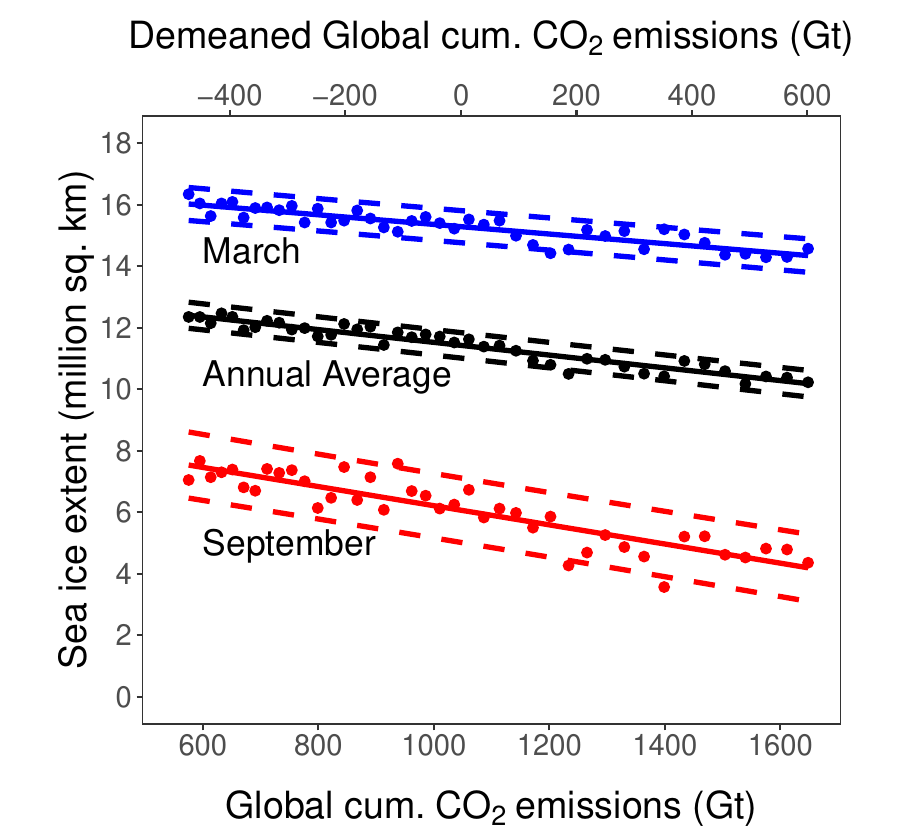}
	\end{center}
	\begin{spacing}{1.0} \footnotesize \noindent Notes: Arctic sea ice extent (in million km$^2$) is shown against global cumulative CO$_2$ emissions in Gt.  March, annual average, and September observations are shown in blue, black, and red, respectively. Dashed lines are  95\% prediction intervals accounting for parameter estimation uncertainty.   Carbon scales are shown in both demeaned (top scale) and non-demeaned (bottom scale) form.   The sample period is 1979-2019.  See text for details.
	\end{spacing}
\end{figure}

Here  observed-data Arctic  sea ice  sensitivity is examined using an alternative measure of Arctic sea ice, sea ice extent ($SIE$), as $ICE_t$.  Monthly average $SIE$  data from the National Snow and Ice Data Center (NSIDC) are used, January 1979 - December 2019. These data use the NASA team algorithm to convert microwave brightness readings into ice coverage estimates for each grid cell, after which $SIE$ is then calculated as the total area of all cells with at least 15\% coverage. 

The $SIE$ results  are shown in Figure \ref{SIElinear3} and Table \ref{observed2}.  They  parallel the $SIA$ results shown in the main text in Figure \ref{SIAlinear3} and Table \ref{observed}, although, because $SIE$ is invariably a bit higher than $SIA$ by virtue of its construction, the September sea ice NIFA and IFA dates are pushed about 7 or 8 years later.

\begin{table}[tbph]
	\caption{Regressions of sea ice extent on Cumulative CO$_2$ Emissions}
	\label{observed2}
	\begin{center}
		\begin{tabular}{lccc}
			\hline\hline
			\\[-2ex]& \multicolumn{3}{c}{Sea Ice Extent}  \\
			\cline{2-4}
			\\[-2ex]& September & Annual Average & March  \\
			\hline
			\\[-2ex]Intercept (million km$^2$)   & 6.068     & 11.427         & 15.284 \\
			& (.080)    & (.032)         & (.040) \\
			Sensitivity (m$^2$/t CO$_2$) & -3.112    & -2.072         & -1.563 \\
			& (.253)    & (.010)         & (.126) \\
			R$^2$                        & .79       & .92            & .80    \\
			&           &                &        \\
			NIFA CO$_2$ level            & 2675      & 6079           & 10187  \\
			NIFA Year (SSP2-4.5)         & 2042      &                &        \\
			NIFA Year (SSP3-7.0)         & 2039      & 2089           &       \\
			\hline\hline
		\end{tabular}
	\end{center}
	\begin{spacing}{1.0} \footnotesize \noindent  Notes: Estimated coefficients and R$^{2}$'s are shown from regressions of sea ice extent  ($SIE$)  on demeaned Cumulative CO$_2$ Emissions since 1850. Standard errors are in parentheses. Intercepts are reported in million km$^2$. Sensitivity or slope coefficients for cumulative CO$_2$ emissions since 1850 are shown in m$^2$ per ton of CO$_2$ (which is equivalent to thousands km$^2$ per Gt CO$_2$).
		Extrapolated dates of a nearly ice-free Arctic (NIFA) and the associated  CO$_2$ levels are shown assuming the SSP2-4.5 and SSP3-7.0 scenario paths.   The sample period is 1979-2019. See text for details.
	\end{spacing}
\end{table}

\clearpage

\section{Detailed CMIP5 Regression Results}  \label{appendixc}

{\small
	\begin{center}
		\setlength{\tabcolsep}{7pt}
		\begin{longtable}{lccccccr}
			\caption[September sea ice individual model results (Cumulative CO$_2$ Emissions since 1850)]{September sea ice individual model results (Cumulative CO$_2$ Emissions since 1850)} \\
			\label{table2}
			      & \multicolumn{3}{c}{Intercept (mill.   km$^2$)}               & \multicolumn{3}{c}{Slope (m$^2$/t)}                       &            \\
			      \cline{2-4} \cline{5-7}
			Model & $\hat{\alpha}_i$ & s.e. & $P(\hat{\alpha}_i=\hat{\alpha}_0)$ & $\hat{\beta}_i$ & s.e. & $P(\hat{\beta}_i=\hat{\beta}_0)$ & Joint Test \\
			\hline  
			\endhead
			
			\hline \multicolumn{8}{r}{{Continued on next page}} \\ \hline
			\endfoot
			
			\hline \hline
			\endlastfoot
			
			Observed           & 4.587 & 0.066 & ---    & -2.891 & 0.209 & ---    & ---    \\
			\hline
			ACCESS1-0          & 4.849 & 0.097 & 0.028 & -1.941 & 0.307 & 0.013 & 0.005 \\
			ACCESS1-3          & 4.869 & 0.095 & 0.017 & -2.633 & 0.301 & 0.482 & 0.045 \\
			BNU-ESM            & 2.329 & 0.066 & 0.000 & -2.163 & 0.209 & 0.016 & 0.000 \\
			CanESM2\_R1        & 2.416 & 0.085 & 0.000 & -2.422 & 0.271 & 0.174 & 0.000 \\
			CanESM2\_R2        & 2.494 & 0.085 & 0.000 & -2.087 & 0.269 & 0.021 & 0.000 \\
			CanESM2\_R3        & 2.559 & 0.071 & 0.000 & -2.192 & 0.224 & 0.025 & 0.000 \\
			CanESM2\_R4        & 2.429 & 0.079 & 0.000 & -2.763 & 0.252 & 0.695 & 0.000 \\
			CanESM2\_R5        & 2.464 & 0.092 & 0.000 & -2.277 & 0.291 & 0.091 & 0.000 \\
			CCSM4\_R1          & 5.335 & 0.116 & 0.000 & -2.347 & 0.369 & 0.203 & 0.000 \\
			CCSM4\_R2          & 5.660 & 0.083 & 0.000 & -1.423 & 0.263 & 0.000 & 0.000 \\
			CCSM4\_R3          & 5.285 & 0.096 & 0.000 & -1.669 & 0.305 & 0.001 & 0.000 \\
			CCSM4\_R4          & 5.286 & 0.071 & 0.000 & -1.337 & 0.224 & 0.000 & 0.000 \\
			CCSM4\_R5          & 5.657 & 0.080 & 0.000 & -1.848 & 0.252 & 0.002 & 0.000 \\
			CCSM4\_R6          & 5.435 & 0.077 & 0.000 & -2.133 & 0.244 & 0.021 & 0.000 \\
			CESM1-BGC          & 5.323 & 0.083 & 0.000 & -2.913 & 0.264 & 0.950 & 0.000 \\
			CESM1-CAM5         & 5.498 & 0.092 & 0.000 & -2.576 & 0.292 & 0.382 & 0.000 \\
			CESM1-CAM5-1-FV2   & 5.913 & 0.101 & 0.000 & -1.762 & 0.319 & 0.004 & 0.000 \\
			CESM1-WACCM\_R1    & 5.435 & 0.077 & 0.000 & -2.133 & 0.244 & 0.021 & 0.000 \\
			CESM1-WACCM\_R2    & 7.648 & 0.065 & 0.000 & -1.486 & 0.207 & 0.000 & 0.000 \\
			CESM1-WACCM\_R3    & 7.614 & 0.059 & 0.000 & -0.916 & 0.186 & 0.000 & 0.000 \\
			CESM1-WACCM\_R4    & 7.538 & 0.041 & 0.000 & -0.856 & 0.131 & 0.000 & 0.000 \\
			CMCC-CESM          & 7.808 & 0.043 & 0.000 & -1.184 & 0.138 & 0.000 & 0.000 \\
			CMCC-CM            & 8.772 & 0.052 & 0.000 & -2.988 & 0.163 & 0.717 & 0.000 \\
			CMCC-CMS           & 7.392 & 0.039 & 0.000 & -0.998 & 0.123 & 0.000 & 0.000 \\
			CNRM-CM5\_R1       & 4.617 & 0.098 & 0.798 & -2.479 & 0.310 & 0.274 & 0.530 \\
			CNRM-CM5\_R2       & 4.057 & 0.106 & 0.000 & -1.491 & 0.335 & 0.001 & 0.000 \\
			CNRM-CM5\_R4       & 4.324 & 0.105 & 0.038 & -2.328 & 0.334 & 0.157 & 0.044 \\
			CNRM-CM5\_R6       & 4.520 & 0.088 & 0.543 & -2.856 & 0.279 & 0.921 & 0.826 \\
			CNRM-CM5\_R10      & 4.178 & 0.154 & 0.017 & -4.291 & 0.487 & 0.010 & 0.002 \\
			CSIRO-Mk3-6-0\_R1  & 9.281 & 0.027 & 0.000 & -1.069 & 0.087 & 0.000 & 0.000 \\
			CSIRO-Mk3-6-0\_R2  & 9.417 & 0.034 & 0.000 & -0.779 & 0.107 & 0.000 & 0.000 \\
			CSIRO-Mk3-6-0\_R3  & 9.293 & 0.026 & 0.000 & -1.125 & 0.084 & 0.000 & 0.000 \\
			CSIRO-Mk3-6-0\_R4  & 9.695 & 0.033 & 0.000 & -0.876 & 0.106 & 0.000 & 0.000 \\
			CSIRO-Mk3-6-0\_R5  & 9.645 & 0.026 & 0.000 & -1.139 & 0.083 & 0.000 & 0.000 \\
			CSIRO-Mk3-6-0\_R6  & 9.728 & 0.029 & 0.000 & -0.979 & 0.092 & 0.000 & 0.000 \\
			CSIRO-Mk3-6-0\_R7  & 9.405 & 0.036 & 0.000 & -1.426 & 0.116 & 0.000 & 0.000 \\
			CSIRO-Mk3-6-0\_R8  & 9.506 & 0.036 & 0.000 & -1.478 & 0.113 & 0.000 & 0.000 \\
			CSIRO-Mk3-6-0\_R9  & 9.375 & 0.035 & 0.000 & -0.364 & 0.112 & 0.000 & 0.000 \\
			CSIRO-Mk3-6-0\_R10 & 9.525 & 0.045 & 0.000 & -1.017 & 0.142 & 0.000 & 0.000 \\
			EC-EARTH\_R1       & 6.505 & 0.046 & 0.000 & -1.687 & 0.146 & 0.000 & 0.000 \\
			EC-EARTH\_R2       & 6.658 & 0.056 & 0.000 & -1.474 & 0.177 & 0.000 & 0.000 \\
			EC-EARTH\_R3       & 6.748 & 0.070 & 0.000 & -1.900 & 0.221 & 0.002 & 0.000 \\
			EC-EARTH\_R4       & 9.695 & 0.033 & 0.000 & -0.876 & 0.106 & 0.000 & 0.000 \\
			EC-EARTH\_R5       & 7.654 & 0.123 & 0.000 & 2.729  & 0.391 & 0.000 & 0.000 \\
			EC-EARTH\_R6       & 6.681 & 0.071 & 0.000 & -1.046 & 0.224 & 0.000 & 0.000 \\
			EC-EARTH\_R8       & 6.754 & 0.089 & 0.000 & -1.278 & 0.283 & 0.000 & 0.000 \\
			EC-EARTH\_R9       & 6.686 & 0.061 & 0.000 & -1.300 & 0.194 & 0.000 & 0.000 \\
			EC-EARTH\_R10      & 6.656 & 0.052 & 0.000 & -1.667 & 0.165 & 0.000 & 0.000 \\
			EC-EARTH\_R12      & 6.541 & 0.069 & 0.000 & -1.833 & 0.218 & 0.001 & 0.000 \\
			EC-EARTH\_R14      & 6.903 & 0.062 & 0.000 & -0.976 & 0.195 & 0.000 & 0.000 \\
			FGOALS-g2          & 6.803 & 0.061 & 0.000 & -0.574 & 0.193 & 0.000 & 0.000 \\
			FIO-ESM\_R1        & 4.969 & 0.073 & 0.000 & -0.988 & 0.233 & 0.000 & 0.000 \\
			FIO-ESM\_R2        & 5.320 & 0.055 & 0.000 & -1.182 & 0.174 & 0.000 & 0.000 \\
			FIO-ESM\_R3        & 5.007 & 0.076 & 0.000 & -1.222 & 0.243 & 0.000 & 0.000 \\
			GFDL-CM3           & 5.343 & 0.095 & 0.000 & -3.547 & 0.303 & 0.079 & 0.000 \\
			GFDL-ESM2G         & 6.820 & 0.077 & 0.000 & -1.791 & 0.245 & 0.001 & 0.000 \\
			GFDL-ESM2M         & 5.168 & 0.088 & 0.000 & -0.475 & 0.279 & 0.000 & 0.000 \\
			GISS-E2-H          & 1.399 & 0.095 & 0.000 & -2.093 & 0.301 & 0.032 & 0.000 \\
			GISS-E2-H-CC       & 1.588 & 0.104 & 0.000 & -1.286 & 0.331 & 0.000 & 0.000 \\
			GISS-E2-R\_R1      & 2.429 & 0.091 & 0.000 & -2.037 & 0.288 & 0.019 & 0.000 \\
			GISS-E2-R\_R2      & 2.445 & 0.084 & 0.000 & -2.478 & 0.268 & 0.228 & 0.000 \\
			HadGEM2-ES\_R1     & 3.590 & 0.081 & 0.000 & -3.054 & 0.257 & 0.624 & 0.000 \\
			HadGEM2-ES\_R2     & 3.946 & 0.080 & 0.000 & -2.079 & 0.255 & 0.016 & 0.000 \\
			HadGEM2-ES\_R3     & 3.409 & 0.093 & 0.000 & -2.271 & 0.294 & 0.090 & 0.000 \\
			HadGEM2-ES\_R4     & 3.617 & 0.081 & 0.000 & -3.359 & 0.258 & 0.164 & 0.000 \\
			IPSL-CM5A-LR\_R1   & 5.902 & 0.074 & 0.000 & -2.205 & 0.235 & 0.032 & 0.000 \\
			IPSL-CM5A-LR\_R2   & 5.238 & 0.083 & 0.000 & -1.503 & 0.264 & 0.000 & 0.000 \\
			IPSL-CM5A-LR\_R3   & 6.062 & 0.066 & 0.000 & -2.083 & 0.209 & 0.008 & 0.000 \\
			IPSL-CM5A-LR\_R4   & 5.867 & 0.069 & 0.000 & -1.818 & 0.218 & 0.001 & 0.000 \\
			IPSL-CM5A-MR       & 4.226 & 0.083 & 0.001 & -2.269 & 0.262 & 0.067 & 0.001 \\
			IPSL-CM5B-LR       & 8.032 & 0.064 & 0.000 & -1.117 & 0.204 & 0.000 & 0.000 \\
			MIROC5\_R1         & 5.820 & 0.086 & 0.000 & -2.525 & 0.271 & 0.289 & 0.000 \\
			MIROC5\_R2         & 5.880 & 0.054 & 0.000 & -2.086 & 0.170 & 0.004 & 0.000 \\
			MIROC5\_R3         & 5.888 & 0.065 & 0.000 & -1.864 & 0.206 & 0.001 & 0.000 \\
			MIROC-ESM          & 4.719 & 0.060 & 0.141 & -1.815 & 0.189 & 0.000 & 0.000 \\
			MIROC-ESM-CHEM     & 5.157 & 0.096 & 0.000 & -2.751 & 0.304 & 0.704 & 0.000 \\
			MPI-ESM-LR\_R1     & 4.840 & 0.071 & 0.011 & -1.955 & 0.224 & 0.003 & 0.001 \\
			MPI-ESM-LR\_R2     & 4.915 & 0.094 & 0.006 & -1.477 & 0.299 & 0.000 & 0.000 \\
			MPI-ESM-LR\_R3     & 4.835 & 0.054 & 0.005 & -1.293 & 0.171 & 0.000 & 0.000 \\
			MPI-ESM-MR         & 4.890 & 0.068 & 0.002 & -1.755 & 0.216 & 0.000 & 0.000 \\
			MRI-CGCM3          & 5.076 & 0.149 & 0.004 & -1.330 & 0.474 & 0.003 & 0.000 \\
			MRI-ESM1           & 5.015 & 0.162 & 0.017 & -2.437 & 0.513 & 0.415 & 0.041 \\
			NorESM1-M          & 6.670 & 0.053 & 0.000 & -1.498 & 0.167 & 0.000 & 0.000 \\
			NorESM1-ME         & 7.306 & 0.064 & 0.000 & -0.991 & 0.202 & 0.000 & 0.000 \\
			\hline
		\end{longtable}
		\begin{flushleft}
			{\footnotesize \noindent Notes: The sea ice sensitivity regression intercept and slope estimated from  the observed September data (1979-2019) are shown on the first line of the table.  Sea ice sensitivity regression intercepts and slopes estimated from CMIP5 models are shown in the $\hat{\alpha}_i$ and $\hat{\beta}_i$ columns respectively along with standard errors for each estimate.  The columns $P(\hat{\alpha}_i=\hat{\alpha}_0)$ and $P(\hat{\beta}_i=\hat{\beta}_0)$ show the p-values for the tests $H_0: a=0$ and $H_0: b=0$ respectively. The $a$ and $b$ coefficients are estimated from the regression $SIA_t= \alpha_i + \beta_i CO2_t + aO_t + bO_tCO2_t + \varepsilon_t$ performed on model and observed pooled data, where $O_t$ is a dummy variable for observed data. The last column shows the p-value of the joint test $H_0: (a,b)=(0,0)$. }
		\end{flushleft}
	\end{center}
}

\clearpage
\bibliographystyle{Diebold}
\addcontentsline{toc}{section}{References}
\bibliography{Bibliography}

\end{document}